\documentclass[apj]{emulateapj}
\usepackage{url}

\def\degpnt{^{\circ}\kern-1.7mm.\kern+.35mm}
\def\arcpnt{"\kern-1.7mm.\kern+.35mm}
\def\minpnt{'\kern-1.0mm.\kern+.30mm}

\slugcomment{Accepted to ApJ}

\shorttitle{Extreme SN Hosts}
\shortauthors{Neill, et al.}

\begin{document}


\title{The Extreme Hosts of Extreme Supernovae}


\author{James~D.~Neill\altaffilmark{1}, Mark~Sullivan\altaffilmark{2},
	Avishay~Gal-Yam\altaffilmark{3}, Robert~Quimby\altaffilmark{1}, 
	Eran~Ofek\altaffilmark{1}, Ted~K.~Wyder\altaffilmark{1}, 
	D.~Andrew~Howell\altaffilmark{4}, Peter~Nugent\altaffilmark{5},
	Mark~Seibert\altaffilmark{6}, D.~Christopher~Martin\altaffilmark{1},
	Roderik~Overzier\altaffilmark{7}, 
	Tom A. Barlow\altaffilmark{1}, Karl~Foster\altaffilmark{1},
	Peter~G.~Friedman\altaffilmark{1}, Patrick~Morrissey\altaffilmark{1},
	Susan~G.~Neff\altaffilmark{8}, David~Schiminovich\altaffilmark{9},
	Luciana~Bianchi\altaffilmark{10}, Jos\'e~Donas\altaffilmark{11},
	Timothy~M.~Heckman\altaffilmark{12}, Young-Wook~Lee\altaffilmark{13},
	Barry~F.~Madore\altaffilmark{6}, Bruno~Milliard\altaffilmark{11},
	R.~Michael~Rich\altaffilmark{14}, and Alex~S.~Szalay\altaffilmark{12}
}

\altaffiltext{1}{California Institute of Technology, 1200 E. California Blvd., Pasadena, CA 91125, USA}
\altaffiltext{2}{University of Oxford, Denys Wilkinson Building,
Keble Road, Oxford, OX1 3RH, UK}
\altaffiltext{3}{Department of Particle Physics and Astrophysics, Faculty of Physics, Weizmann Institute of Science, 76100 Rehovot, Israel}
\altaffiltext{4}{Las Cumbres Observatory Global Telescope Network, 6740 Cortona Dr., Suite 102, Goleta, CA 93117, USA}
\altaffiltext{5}{Lawrence Berkeley National Laboratory, MS 50F-1650, 1
Cyclotron Road, Berkeley, CA, 94720-8139, USA}
\altaffiltext{6}{The Observatories of the Carnegie Institute of Washington, 813 Santa Barbara Street, Pasadena, CA, 91101, USA}
\altaffiltext{7}{Max-Planck-Institut f\"ur Astrophysik, Karl-Schwarzschild-Str. 1, D-85748 Garching, Germany}
\altaffiltext{8}{Laboratory for Astronomy and Solar Physics, NASA Goddard
Space Flight Center, Greenbelt, MD, 20771, USA}
\altaffiltext{9}{Department of Astronomy, Columbia University, New
York, NY 10027, USA}
\altaffiltext{10}{Center for Astrophysical Sciences, The Johns Hopkins  
University, 3400 N. Charles St., Baltimore, MD, 21218, USA}
\altaffiltext{11}{Laboratoire d'Astrophysique de Marseille, BP 8,  
Traverse du Siphon, 13376 Marseille Cedex 12, France}
\altaffiltext{12}{Department of Physics and Astronomy, The Johns  
Hopkins University, Homewood Campus, Baltimore, MD 21218, USA}
\altaffiltext{13}{Center for Space Astrophysics, Yonsei University,  
Seoul 120-749, Korea}
\altaffiltext{14}{Department of Physics and Astronomy, University of  
California, Los Angeles, CA 90095, USA}


\begin{abstract}

We use {\it GALEX} ultraviolet (UV) and optical integrated photometry of
the hosts of seventeen luminous supernovae (LSNe, having peak $M_V<-21$)
and compare them to a sample of $26,000$ galaxies from a cross-match
between the SDSS DR4 spectral catalog and {\it GALEX} interim release 1.1.
We place the LSNe hosts on the galaxy $NUV-r$ versus $M_r$ color magnitude
diagram (CMD) with the larger sample to illustrate how extreme they are.
The LSN hosts appear to favor low-density regions of the galaxy CMD falling
on the blue edge of the blue cloud toward the low luminosity end.  From the
UV-optical photometry, we estimate the star formation history of the LSN
hosts.  The hosts have moderately low star formation rates (SFRs) and low
stellar masses ($M_*$) resulting in high specific star formation rates
(sSFR).  Compared with the larger sample, the LSN hosts occupy low-density
regions of a diagram plotting $sSFR$ versus $M_*$ in the area having higher
$sSFR$ and lower $M_*$.  This preference for low $M_*$, high $sSFR$ hosts
implies the LSNe are produced by an effect having to do with their local
environment.  The correlation of mass with metallicity suggests that perhaps
wind-driven mass loss is the factor that prevents LSNe from arising in
higher-mass, higher-metallicity hosts.  The massive progenitors of the LSNe
($>100 M_{\odot}$), by appearing in low-SFR hosts, are potential tests for
theories of the initial mass function that limit the maximum mass of a star
based on the SFR.

\end{abstract}




\section{Introduction\label{sec_intro}}

Two extremely luminous core-collapse (CC) supernovae (SNe) were recently
discovered with faint or non-detected hosts, one at low redshift
\citep[SN2005ap, $M_{Bol,Peak}=-22.7$ at $z=0.283$,][]{Quimby:07:L99} and
one with {\it HST} at a higher redshift \citep[SCP 06F6,
$M_{Bol,Peak}=-22.1$ at $z=1.189$,][]{Barbary:09:1358,Quimby:09}.  Recent
wide-area surveys \citep[e.g.,][]{Quimby:06:13,  Rau:09:1334, Law:09:1395,
Drake:09:870} have discovered similar objects revealing a new class of
extremely luminous CC SN \citep{Quimby:07:L99, Ofek:07:L13, Smith:08:467,
Yuan:08:1, Miller:09:1303, Gezari:09:1313, Yam:09:624, Quimby:10:1,
Drake:10:L127, Pastorello:10:L16} that were missed in earlier host-targeted
surveys due to their preference for faint hosts \citep{Young:10:70,
Quimby:09, Yam:09:624}.  It is not uncommon that other types of extremely
luminous SNe are found in low-luminosity hosts \citep[see,
e.g.,][]{Kozlowski:10:1624}, for example many extreme SNe IIn (see
\S\ref{sec_data}) seem to prefer dwarf hosts \citep{Richardson:02:745,
Smith:08:467, Miller:09:1303}, but not always \citep{Smith:07:1116}.

The preference these extremely luminous SNe (LSNe) have for low-mass and
presumably low-metallicity hosts implies a factor in their production
specific to the host galaxy.  It is thus important to begin to quantify the
local host environments of the LSNe.  The link between galaxy mass and
metallicity demonstrated in \citet{Tremonti:04:898} and the LSN preference
for low-mass hosts could imply that metallicity has an influence on the
stellar initial mass function (IMF).  This preference for low-mass hosts
could also be a natural consequence of an increase in the efficiency of
metal-line driven stellar winds that lower the final masses of these same
objects in larger galaxies and consequently produce lower-luminosity
explosions \citep[see, e.g.,][]{Arcavi:10:777}.  The extreme masses of the
LSNe progenitors in low-mass galaxies offers the opportunity to test models
of the IMF that posit distributions limiting the most massive star based
on available star-forming gas \citep{Altenburg:07:1550, Weidner:10:275}.

Three scenarios have been explored that can produce LSNe: the large
production of radioactive fuel produced from the thermonuclear burning of a
massive oxygen core that is instigated by a pulsational electron-positron
pair instability \citep[PISNe][]{Barkat:67:379, Bond:84:825, Heger:02:532,
Waldman:08:1103}, the interaction of the outburst with a dense
circumstellar envelope either left over from progenitor formation
\citep{Metzger:10:284} or produced by late time mass-loss 
\citep{Yam:07:372, Smith:07:1116, Smith:08:467}, and the energy
from a rapidly-rotating magentar formed in the collapse of the LSN
progenitor star \citep{Kasen:10:245, Woosley:10:L204}.  The first two
scenarios naturally imply extreme masses for the progenitors of LSNe, in
some cases in excess of 150~$M_{\odot}$ \citep{Yam:09:624}.

\begin{figure}
\includegraphics[angle=90.,scale=0.425]{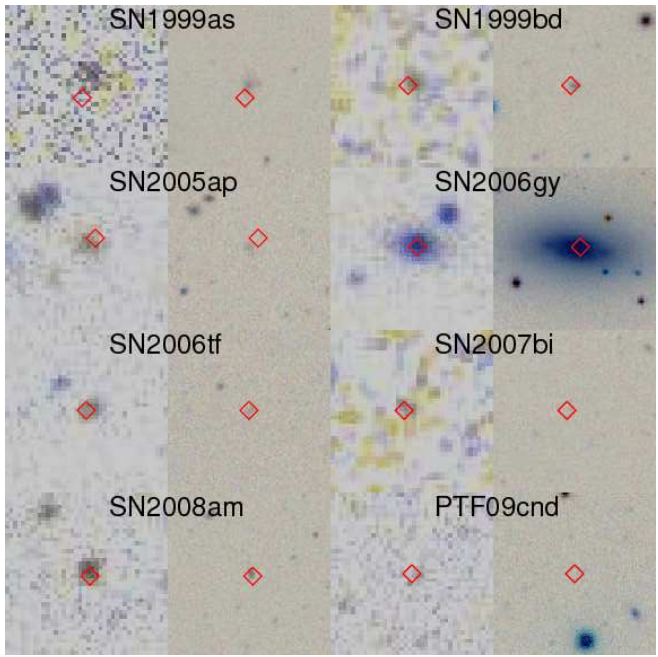}
\caption{{\it GALEX} FUV/NUV psuedo-three color images (left panels) and SDSS 
images (right panels) of the detected hosts of the extreme supernovae.  Each
panel is one arcminute across.  The red diamond marks the location of the 
supernova.  The real host of SN2005ap is blended with a nearby galaxy
\citep[see][]{Quimby:07:L99}, so the {\it GALEX} images were used to 
determine an upper limit.}
\label{fig_sne}
\end{figure}

Theoretical treatments of the magnetar scenario predict the basic form of
observed LSN light curves \citep{Kasen:10:245, Woosley:10:L204}, but have
yet to provide predictions of detailed spectral features, unlike the
pair-production theory which matches the light curves
\citep{Scannapieco:05:1031} and predicts the production of Fe-group
elements that have been observed \citep{Yam:09:624}.  Without these
predictions it is yet unclear how many, if any, of the LSNe are
magnetar-powered.  The same wind-driven mass loss (WDML) and hence
correlation with host metallicity and mass could operate in this scenario,
if it turns out that magnetar progenitors are also highly massive.  The
mass of magnetar progenitors is currently being debated in the literature
\citep{Klose:04:L13, Gaensler:05:L95, Muno:06:L41, Davies:09:844} and range
from $\sim20$ to $\geq 50 M_{\odot}$.  Until we can confirm that magnetars
have actually powered a LSN, it is safer to make the assumption that the
LSNe arise from very massive progenitors.

From a galaxy evolution standpoint, the formation of massive stars in
low-mass dwarf galaxies implies the very high-density star formation
typically found in UV-luminous galaxies \citep[UVLGs,][]{Hoopes:07:441} and
local Lyman-break analogs \citep[LBAs,][]{Overzier:09:203}.  The LSN hosts
may delimit the low-luminosity, low-mass, low-metallicity range of these
extreme compact starbursting objects, that have not yet made it into
current UVLG and LBA samples.  It is thus important to compare the
properties of the LSN hosts with those of the UVLGs and LBAs.

We begin the examination of LSNe local environments by comparing the hosts
of seventeen of the most luminous SNe on record with a sample of $26,000$
galaxies from a cross-match \citep{Wyder:07:293} between the SDSS
spectroscopy catalog and the Galaxy Evolution Explorer \citep[{\it
GALEX},][]{Martin:05:L1} IR1.1 catalogs.  We use UV and optical photometry
of the hosts to fit star formation history (SFH) models and estimate their
luminosity-weighted age ($\langle Age\rangle_L$), stellar mass ($M_*$), and
current star formation rate (SFR).  We compare the distributions of the LSN
hosts with the larger sample on the galaxy $NUV-r$ versus $M_r$ color
magnitude diagram (CMD) and a diagram plotting specific star formation rate
($sSFR = SFR/M_*$) versus $M_*$ to demonstrate their extreme nature and to
explore other relationships between the LSN subtypes and their host
properties.  We also compare the SFR of the LSN hosts with models that
relate SFR to the IMF \citep{Altenburg:07:1550} to estimate the probability
of producing the high-mass stars capable of producing LSNe.

\section{Data}
\label{sec_data}

Our initial sample of seventeen LSNe consists of all those discovered to
date with $M_V <-21$ as derived from modern (post 1990) photometry
\citep{Richardson:02:745, Quimby:07:L99, Ofek:07:L13, Smith:08:467,
Yuan:08:1, Miller:09:1303, Gezari:09:1313, Yam:09:624, Drake:10:L127,
Pastorello:10:L16}.  Our
sample includes the LSNe produced by the interaction of the explosion
ejecta and the surrounding circum-stellar matter.  This subgroup is
characterized by narrow emission lines in their spectra and are called
Type~IIn-lum.  Two of our sample (SN1999as and SN2007bi) have been
determined to be PISNe by their light-curves and by showing Fe-group
elements in their spectra \citep{Yam:09:624} and are labeled Ic-PP.  The
other class of LSN studied here we label Type Ipec, which denotes the lack
of H emission in their spectra and their peculiar properties (spectra and
lightcurve) when compared with any other SN type \citep{Quimby:09}.  These
are possibly pulsational PISNe as well \citep{Quimby:09}, but their
spectra and light-curves do not allow a conclusive classification.

For a comparison sample we use the cross-match between the spectral sample
of SDSS DR4 and the {\it GALEX} G1 interim release IR1.1 catalog presented
in \citet{Wyder:07:293}.  The basic sample criteria limit the apparent SDSS
r-band magnitude of this sample to 17.6 and limit the redshift to $0.01 < z
< 0.25$.  The NUV completeness for blue galaxies on the faint end is
$\sim90$\% for this sample, while the NUV faint limit does lead to higher
incompleteness for faint red galaxies.  For more details on the sample
selection, see \S 2.2 of \citet{Wyder:07:293}.

\begin{deluxetable*}{llrlrllll}[ht]
	\tablewidth{0in}
	\tabletypesize{\scriptsize}
	\tablecaption{Luminous SNe Host Photometry\label{tab_sne}}
	\tablehead{ & & & & \colhead{Exptime} & \multicolumn{2}{c}{Observed} &
	\multicolumn{2}{c}{K-corrected ($z=0.1$)}\\
	\colhead{SN} & \colhead{Type} & \colhead{z} & \colhead{HOST} &
	\colhead{NUV s} & \colhead{NUV Mag} & \colhead{r Mag\tablenotemark{1}} &
	\colhead{NUV Mag} & \colhead{r Mag\tablenotemark{1}}
	}
\startdata
1995av   & IIn-lum\tablenotemark{2}  & 0.300 & A020136.0+033855.0  &   440 & $>$22.92 & $>$22.80\tablenotemark{3} & $>$22.62 & $>$22.45\tablenotemark{3}\\
1997cy   & IIn-lum  & 0.063 & A043255.1$-$614300.0  &  1811 & 19.95$\pm$0.04 & 19.72$\pm$0.20\tablenotemark{4} & 19.95$\pm$0.04 & 19.72$\pm$0.20\tablenotemark{4}\\
1999as   & Ic-PP  & 0.127 & SDSS J091630.79+133906.1 &   358 &  20.42$\pm$0.13 & 19.20$\pm$0.08 & 20.42$\pm$0.13 & 19.20$\pm$0.08\\
1999bd   & IIn-lum & 0.151 & SDSS J093029.10+162607.1  &   247 &  21.88$\pm$0.27 & 19.81$\pm$0.14 & 21.88$\pm$0.27 & 19.81$\pm$0.14\\
2000ei   & IIn-lum\tablenotemark{2}  & 0.600 & SDSS J041707.06+054551.8  &
2147 & $>$21.73 & 22.75$\pm$0.71 & $>$20.93 & 21.60$\pm$0.71 \\
2005ap   & Ipec  & 0.283 & A130113.1+274334.4  &  4123 & $>$23.49 & 23.71$\pm$0.25\tablenotemark{5} & $>$23.19 & 23.36$\pm$0.25\tablenotemark{5}\\
2006gy   & IIn-lum & 0.019 & NGC1260       & 21783 & 17.89$\pm$0.02 & 12.08$\pm$0.01 & 17.89$\pm$0.02 & 12.08$\pm$0.01\\
2006tf   & IIn-lum & 0.074 & SDSS J124615.80+112555.5  &  3298 & 21.43$\pm$0.07 & 20.75$\pm$0.34 & 21.43$\pm$0.07 & 20.75$\pm$0.34\\
2007bi   & Ic-PP   & 0.128 & SDSS J131920.14+085543.7  &   202 & 21.37$\pm$0.27 & 22.41$\pm$0.76 & 21.37$\pm$0.27 & 22.41$\pm$0.76\\
2008am   & IIn-lum & 0.234 & SDSS J122836.31+153449.5  &  2167 & 21.02$\pm$0.06 & 19.93$\pm$0.14 & 20.82$\pm$0.06 & 19.73$\pm$0.14\\
2008es   & IIn-lum & 0.202 & A115649.0+542725.0  &     0 & \nodata & $>$21.71 & \nodata & \nodata\\
2008fz   & IIn-lum & 0.133 & A231616.5+114248.5  &   192 & 21.01$\pm$0.23 & $>$21.33\tablenotemark{3} & 21.01$\pm$0.23 & $>$21.33\tablenotemark{3}\\
SCP06F6  & Ipec    & 1.189 & A143227.4+333224.8  & 69291 & $>$25.98 & $>$22.80 & $>$24.88 & $>$20.80\\
PTF09atu & Ipec    & 0.501 & A163024.5+233825.0  &   543 & $>$22.93 & $>$22.61 & $>$22.23 & $>$21.61\\
PTF09cnd & Ipec    & 0.258 & A161209.0+512914.5  &  1951 & 23.19$\pm$0.22 & $>$21.67 & 22.89$\pm$0.22 & $>$21.37\\
2009jh & Ipec    & 0.349 & A144910.1+292511.4  &  3249 & $>$24.34 & $>$21.77 & $>$23.94 & $>$21.37\\
2010gx & Ipec    & 0.230 & SDSS J112546.72$-$084942.0  &   423 & $>$23.09 & 22.42$\pm$0.24 & $>$22.89 & 22.22$\pm$0.24
\enddata
\tablenotetext{1}{SDSS $r$-band, unless otherwise noted}
\tablenotetext{2}{Classification uncertain}
\tablenotetext{3}{Derived from DeepSky photometry \citep{Nugent:09:419}}
\tablenotetext{4}{Derived from photometry presented in \citet{Germany:00:320}}
\tablenotetext{5}{Derived from photometry presented in \citet{Adami:06:1159}}
\end{deluxetable*}

The apparent magnitudes of the LSNe hosts are much fainter than this larger
sample.  In addition, seven of the seventeen LSN hosts are outside the
sample redshift range.  Our goal in this initial study is not to measure
the relative frequency of LSN hosts in the local universe.  Our goal is,
instead, to place the LSN hosts in a galaxy evolution context as mapped out
on the galaxy CMD using a well-measured local sample.  We also aim to
illustrate that the LSNe are useful for selecting active dwarf galaxies
that would ordinarily go undiscovered.

To characterize the hosts of the LSNe, we take advantage of the close
correlation between UV luminosity and SFR \citep{Treyer:07:256,
Salim:07:267, Martin:07:415}.  We use archival {\it GALEX}
\citep{Martin:05:L1} $FUV$ ($\lambda_c = 1539$\AA, $\Delta \lambda =
442$\AA) and $NUV$ ($\lambda_c = 2316$\AA, $\Delta \lambda = 1060$\AA)
images and coadd them together to obtain the deepest image possible of the
LSNe hosts.  We add optical photometry to the spectral energy distribution
(SED) characterization where available.  Our primary source for optical
imaging of the hosts is the SDSS \citep{York:00:1579} Data Release
7\footnote{\url{http://www.sdss.org/dr7/}}.  We do not use SDSS catalog
photometry, but instead measure the images ourselves, allowing us to match
the apertures in each waveband.  We supplement our SDSS image photometry
with measurements presented in \citet{Germany:00:320} for SN1997cy, deep
photometry of the Coma Cluster by \citet{Adami:06:1159} for SN2005ap, and
images from the DeepSky
Survey\footnote{\url{http://supernova.lbl.gov/~nugent/deepsky.html}}
\citep{Nugent:09:419} for SN1995av and SN2008fz.  Matched apertures are
used to characterize the host SED and to derive magnitudes or detection
limits in each waveband.  A selection of images focussing on the detected
hosts is presented in Figure~\ref{fig_sne}.  

Table~\ref{tab_sne} presents the basic data for the LSNe: IAU designation,
type, redshift, and host name, followed by the $NUV$ exposure time and
observed $NUV$ and $r$-band magnitudes of the host galaxies.  The host
names beginning with `A' denote an anonymous galaxy with the rest of the
name specifying the J2000 position.  In the cases where the host is not
detected, the position is of the LSN.  A K-correction is made
(see \S~\ref{sec_kcor}) and the K-corrected $NUV$ and $r$-band values form
the final two columns of the table.  All the LSNe but SN2006gy appear in
anonymous or SDSS galaxies (see \S\ref{sec_ssfr_vs_mstar}).  We have
detections in the $NUV$ for nine of the seventeen hosts and upper limits
for all but one of the remaining hosts.  SN2008es is near an UV-bright star
preventing {\it GALEX} observations, therefore no $NUV$ upper limit could
be derived.  For the SDSS $r$-band, we have detections for eight of the
seventeen hosts and upper limits for five.  The host of SN1997cy was
measured by \citet{Germany:00:320} with $V$ and $R$-band imaging.  The host
of SN2005ap was detected in the broad-band ($B$ and $V$-band) catalog
presented in \citet{Adami:06:1159} which we converted to an approximate
$r$-band magnitude.  We use the DeepSky imaging to place upper limits on
the $r$-band luminosity of the hosts of SN1995av and SN2008fz.  We had to
make assumptions about the spectral energy distributions for hosts that
were not in the SDSS survey to convert them to an approximate $r$-band.  We
increase their photometric uncertainties to reflect this situation.  All
magnitudes have been corrected for foreground extinction
\citep{Schlegel:98:525} using the reddening law of \citet{Cardelli:89:245}.

\tabletypesize{\scriptsize}
\setlength{\tabcolsep}{0.02in}
\begin{deluxetable*}{lllrrrrrrrrrrrr}[ht]
\tablecaption{Luminous SNe Host Derived Properties\label{tab_host_sed}}
\tablewidth{0pt}
\tablehead{
 & & &
\colhead{Age$-$} & \colhead{$<$Age$>_L$} & \colhead{Age$+$} &
\colhead{M*$-$} & \colhead{$<$M*$>$} & \colhead{M*$+$} &
\colhead{SFR$-$} & \colhead{$<$SFR$>$} & \colhead{SFR$+$} &
\colhead{sSFR$-$} & \colhead{$<$sSFR$>$} & \colhead{sSFR$+$} \\
\colhead{SN} & \colhead{Type} & \colhead{Host} &
\multicolumn{3}{c}{($\log$ yr)} &
\multicolumn{3}{c}{($\log$ M$_{\odot}$)} &
\multicolumn{3}{c}{($\log$ M$_{\odot}$ yr$^{-1}$)} &
\multicolumn{3}{c}{($\log$ yr$^{-1}$)} 
}
\startdata
1995av           & IIn-lum & A020136.0+033855.0        &   6.78 &   8.07 &   9.99 &   6.75 &   8.11 &   9.69 &   $<$-3.00 &   -0.21 &    0.22 &  $<$-12.00 &   -8.32 &   -6.53 \\
1997cy           & IIn-lum & A043255.1$-$614300.0      &   7.45 &   7.82 &   8.15 &   7.96 &   8.13 &   8.27 &   -0.56 &   -0.34 &   -0.18 &   -8.83 &   -8.47 &   -8.14 \\
1999as           & Ic-PP   & SDSS J091630.79+133906.1  &   7.18 &   7.28 &   7.88 &   8.85 &   8.95 &   9.16 &    0.49 &    0.59 &    0.81 &   -8.68 &   -8.36 &   -8.04 \\
1999bd           & IIn-lum & SDSS J093029.10+162607.1  &   7.58 &   7.88 &   9.31 &   9.09 &   9.33 &   9.86 &   -0.38 &    0.92 &    1.06 &  -10.24 &   -8.42 &   -8.03 \\
2000ei           & IIn-lum & SDSS J041707.06+054551.8  &   6.78 &   6.78 &   9.88 &   8.73 &   9.17 &  11.16 &   $<$-3.00 &    0.79 &    2.07 &  $<$-12.00 &   -8.39 &   -6.65 \\
2005ap           & Ipec    & A130113.1+274334.4        &   6.78 &   9.55 &   9.99 &   7.60 &   9.73 &   9.90 &   $<$-3.00 &   -0.52 &    0.48 &  $<$-12.00 &  -10.25 &   -7.13 \\
2006gy           & IIn-lum & NGC1260                   &   9.68 &   9.97 &  10.10 &  11.22 &  11.35 &  11.44 &   $<$-3.00 & $<$-3.00 & $<$-3.00 & $<$-12.00 & $<$-12.00 & $<$-12.00 \\
2006tf           & IIn-lum & SDSS J124615.80+112555.5  &   6.90 &   7.43 &   8.15 &   7.87 &   8.21 &   8.55 &   -0.51 &   -0.14 &    0.16 &   -9.06 &   -8.35 &   -7.71 \\
2007bi           & Ic-PP   & SDSS J131920.14+085543.7  &   6.78 &   7.11 &   8.95 &   6.56 &   7.07 &   8.03 &   -1.97 &   -1.29 &   -0.48 &  -10.01 &   -8.37 &   -7.04 \\
2008am           & IIn-lum & SDSS J122836.31+153449.5  &   6.95 &   7.54 &   8.32 &   9.12 &   9.41 &   9.64 &    0.60 &    1.07 &    1.20 &   -9.04 &   -8.34 &   -7.92 \\
2008es           & IIn-lum & A115649.0+542725.0        &   7.98 &   8.15 &   8.32 &   1.00 &   5.74 &   8.23 &   $<$-3.00 & -2.58 &   -0.20 &  $<$-12.00 &   -8.32 &   $>$-6.00 \\
2008fz           & IIn-lum & A231616.5+114248.5        &   6.78 &   6.90 &  10.06 &   1.00 &   6.71 &   9.01 &   $<$-3.00 &   -1.67 &   -0.43 &  $<$-12.00 &   -8.38 &   $>$-6.00 \\
SCP06F6          & Ipec    & A143227.4+333224.8        &   6.78 &   8.27 &   9.67 &   1.00 &   9.31 &  12.03 &   $<$-3.00 &    0.74 &    1.20 &  $<$-12.00 &   -8.57 &   $>$-6.00 \\
PTF09atu         & Ipec    & A163024.5+233825.0        &   6.78 &   6.78 &   9.73 &   1.00 &   7.30 &  10.07 &   $<$-3.00 &   -1.09 &    0.50 &  $<$-12.00 &   -8.39 &   $>$-6.00 \\
PTF09cnd         & Ipec    & A161209.0+512914.5        &   6.78 &   7.88 &   9.11 &   7.57 &   8.15 &   9.05 &   -0.80 &   -0.18 &    0.23 &   -9.85 &   -8.33 &   -7.34 \\
2009jh         & Ipec    & A144910.1+292511.4        &   6.78 &   7.75 &   9.97 &   1.00 &   7.40 &   9.76 &   $<$-3.00 &   -0.94 &   -0.14 &  $<$-12.00 &   -8.34 &   $>$-6.00 \\
2010gx         & Ipec    & SDSSJ112546.72$-$084942.0 &   6.78 &   8.19 &   9.18 &   7.67 &   8.38 &   8.95 &   -1.15 &   -0.38 &    0.11 &  -10.10 &   -8.76 &   -7.56 \\
\enddata
\end{deluxetable*}

\subsection{K-correction\label{sec_kcor}}

To account for the range in redshift of the LSN hosts and facilitate
comparison with the Wyder sample, we K-correct each of our SN hosts to the
reference redshift of $z=0.1$ used in \citet{Wyder:07:293}.  Since many of
our SN hosts are difficult to measure, we use the 7-band UV-optical
photometry and redshifts available for the Wyder sample galaxies.  We
derive K-corrections for the SN hosts by producing a separate galaxy CMD
for each SN host, with the Wyder galaxies K-corrected to the redshift of
the SN host.  We then compare the diagram at the host redshift with the
diagram at $z=0.1$ to derive an approximate K-correction.  To perform the
K-correction on the Wyder galaxies, we use the latest version of the
K\_CORRECT program \citep{Blanton:07:734} which incorporates the {\it
GALEX} bandpasses.  We estimate that the corrections can be made in this
way to an accuracy of 0.25 magnitudes.  For hosts with $z<0.2$, no
correction is made.  Columns 8 and 9 of Table~\ref{tab_sne} give the
K-corrected magnitudes and errors (not including the K-correction error).

\subsection{Ages, Masses, and Star Formation Rates\label{sec_sfrs}}

For the comparison sample, we use the k-corrected, extinction-corrected
$NUV$ luminosities to estimate their recent ($\lesssim10^8$ yr) SFR
\citep{Treyer:07:256, Salim:07:267}.  These methods assume the
universal IMF from \citet[see \S2.2]{Kroupa:01:231} with a mass
range of 0.1 to 100 $M_{\odot}$.  The internal extinction is estimated
using the Balmer decrement and stellar masses are estimated using spectral
fitting.  For the details of these calculations see \citet{Wyder:07:293}.

Since we do not have spectra of the LSN host galaxies, we cannot use the
same method to derive masses and SFRs.  Instead, we use SED fitting to
models of star formation history (SFH).  To do this we use the known
redshifts and all available photometry and detection limits of the LSN
hosts to fit a particular SFH, from which we estimate host $\langle
Age\rangle_L$, $M_*$, and $SFR$.  This is accomplished with the ZPEG
program, which uses the PEGASE.2 galaxy evolution code \citep{Fioc:97:950,
Borgne:02:446, Borgne:04:881}.  The $SFR$ is also derived using the
univseral IMF from \citet{Kroupa:01:231}.  The models assume exponentially
declining SFHs with over 100 time steps.  Only models that are consistent
with the redshift of the host (i.e., younger than the univsere at that
redshift) are considered in the fitting of the SED.  The metallicity of a
given model is evolved self-consistently \citep{Fioc:97:950}.  A variety of
dust prescriptions were used producing over 100 different SEDs that were
evolved resulting in a grid of over 10$^4$ models. For details on the dust
prescriptions and SFH models, see \citet[][\S3.2]{Sullivan:06:868} and
\citet[][\S2.4]{Sullivan:10:782}. The results of the fitting are presented
in Table~\ref{tab_host_sed} where we repeat the IAU designation, type and
host name for each LSNe, followed by the ranges and most probable values of
$\langle Age\rangle_L$, $M_*$, $SFR$, and $sSFR$ for each host.

To estimate any systematic difference between these two methods, we use the
NUV luminosities and $NUV-r$ colors to estimate internal extinction
\citep{Treyer:07:256} and derive a SFR for each LSN host.  We find an
error-weighted offset between the two SFR methods for the LSNe hosts of
$\Delta SFR [\log(M_{\odot} yr^{-1})] (SED-UV) = 0.54\pm0.10$.  While this
is smaller than the scatter in the SFRs from Table~\ref{tab_host_sed}, we
might be tempted to apply this offset to align the LSN hosts with the
comparison sample, even though we have no way to do the same for stellar
masses.  Unfortunately, the situation for extreme galaxies is complicated
and estimating SFR for extremely blue ($NUV-r < 1$) galaxies shows a high
scatter \citep[see Figure~10 in][]{Treyer:07:256}.  For this reason, we do not
apply any correction.  In addition, with the large photometric errors of
these faint hosts, it is difficult to assess which method is more reliable.

\section{Results\label{sec_results}}

\begin{figure}
\includegraphics[angle=90., scale=0.40,viewport=24 54 528 608]{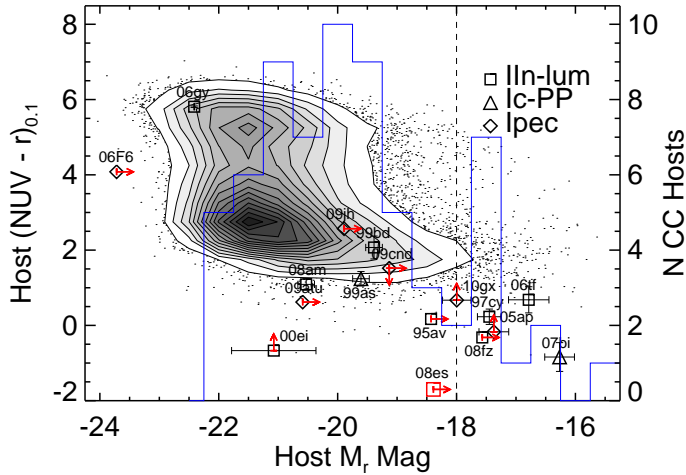}
\caption{Galaxy CMD with hosts of extreme SNe indicated.  The contours
represent the density of galaxies from the {\it GALEX}-SDSS cross match in
\citet{Wyder:07:293} using photometry that is corrected for Milky Way
extinction, K-corrected to a redshift of $z=0.1$ (see text), but
uncorrected for internal extinction.  The arrows indicate limiting
magnitudes derived from existing image data.  The arrows
pointing right limit the host position to a half plane to the right of the
plotted point.  The arrows pointing up limit the color to redward of the
plotted point.  The double arrows for the host of PTF09cnd limit it to a
quarter-plane fainter in $M_r$ and blueward of the plotted point. The blue
histogram plots the CC SN host $M_r$ distribution from \citet{Arcavi:10:777}
refering to the right axis.  The vertical dashed line is the demarcation
between `Giant' and `Dwarf' host galaxies used in that study.
\label{fig_cmd}}
\end{figure}

We plot the seventeen LSN hosts on the galaxy $NUV-r$ versus $M_r$ CMD in
Figure~\ref{fig_cmd}.  The contours represent the galaxy density of the
$\sim$26,000 galaxies from \citet{Wyder:07:293} in the diagram in 0.5 by
0.5 magnitude bins in color and luminosity with the darkest level at a
density of 1056 galaxies per bin and the lightest level at 132 galaxies per
bin.  Below the lowest contour density, the galaxies are plotted
individually as small dots.  We code the symbol for each SN by type: square
for IIn-lum, diamond for Ipec, and triangle for Ic-PP.  For comparison, we
plot the distribution of $M_r$ host magnitudes for the CC SN sample from
\citet{Arcavi:10:777} which refers to the right axis of the figure.  Since
this distribution is derived from an areal survey, it should reflect an
unbiased sampling of the parent distribution of CC SN host magnitudes.

For SN hosts for which we only have an upper limit, we use (red) arrows to
indicate what the allowable range of magnitudes or colors is restricted to.
These arrows indicate that with upper limits in both $NUV$ and $r$, we can
only restrict a half-plane in this diagram.  For these limits, the symbol
is plotted at the position calculated from the color and luminosity of the
limiting magnitudes.  Four of the limiting cases are unusual.  For
PTF09cnd, we have a detection in the $NUV$, but no detection in the $r$
(see Figure~\ref{fig_sne}).  This allows us to limit the color and
luminosity to a quarter-plane blue-ward of and fainter in $M_r$ than the
plotted symbol.  For SN2008es, we have no $NUV$ data (see
\S~\ref{sec_data}), but our $r$-band limit still restricts the allowable
$M_r$.  Therefore, we place the type symbol (red) and identification along
the bottom edge of the diagram.  For SN2005ap and SN2010gx, the optical
detections allow us to measure $M_r$, but the limits in $NUV$ only allow us
to limit the $NUV-r$ color to be red-ward of (above) the symbol.

We now examine the physical conditions within the host galaxies and plot
the seventeen LSNe hosts on a diagram of $sSFR$ versus $M_*$ in
Figure~\ref{fig_ssfr}.  The contours for the larger sample in this case are
derived in two-dimensional bins of 0.2 dex wide in $M_*$ and 0.1 dex wide in
$sSFR$.  The darkest contour represents a density of 282 galaxies per bin
and the lightest contour 17 galaxies per bin, with individual galaxies
plotted below this density.  The symbol coding for the LSNe hosts is the
same as in Figure~\ref{fig_cmd}.  For comparison, we plot the Large
Magellanic Cloud as the solid circle, using the mass range reported in
\citet{Westerlund:97} and the current SFR from \citet{Harris:09:1243}.
Since we are comparing individual galaxies to the larger sample, we
do not apply a volume correction to the larger sample.  This
accounts for the difference between our Figure~\ref{fig_ssfr} and Figure~26
from \citet{Wyder:07:293}.

\section{Discussion\label{sec_discuss}}

Figures~\ref{fig_cmd} and \ref{fig_ssfr} together support the notion that
there is an environmental factor in the production of LSNe.  Their
distribution in the figures tends toward extreme regions of low luminosity,
blue $NUV-r$, low mass, but high sSFR in spite of having low SFR.
Undoubtedly there is incompleteness in the larger sample, but it is hard to
imagine a scenario where incompleteness dominates the distribution of the
LSNe, given that many fainter SNe are found in more luminous galaxies
\citep{Arcavi:10:777}, and that these more luminous galaxies were
preferentially surveyed for decades before areal SN searches were feasible.
The incompleteness in the larger sample limits our ability to say just how
rare the LSN hosts are, but with upcoming deeper, wide-field, multi-band
surveys, the incompleteness limits will be pushed fainter and allow us to
measure their volumetric density.  For now, the presence of a LSN within a
low-luminosity host indicates that the host is undergoing an episode of
active, high-density, high-mass star formation.  The extreme luminosity of
LSNe, allowing them to be detected to high redshift, make LSNe guides for
our exploration of star formation in dwarf galaxies over a range of
redshifts.

\subsection{The Galaxy CMD\label{sec_gcmd}}

Looking at Figure~\ref{fig_cmd} in detail shows that our sample is not
large enough to distinguish the LSNe host $M_r$ distribution from the full
CC host $M_r$ distribution shown in \citet{Arcavi:10:777}.  If we divide
the sample at the central minimum in the Arcavi distribution ($M_r = -19$)
and exclude SN2006gy (see \S\ref{sec_ssfr_vs_mstar}) and SCP06F6 (because
of its high redshift and very uncertain K-correction), we count seven hosts
brighter than this value and eight hosts fainter.  Three of the brighter
hosts are upper limits and some or all could be counted in the fainter
group.  Clearly nothing conclusive can be derived from this, but it begs
for deeper photometry and larger samples to see if the LSNe hosts are drawn
from a different, lower luminosity parent population than the other CC SN
hosts.

We also do not see a definitive separation between IIn-lum hosts and Ipec
or Ic-PP hosts.  The SNe IIn-lum must have had some level of mass loss in
order to produce the narrow lines in their spectra from circum-stellar
interaction.  This could have resulted from WDML or from binary interaction
or a combination of the two.  If IIn-lum hosts are systematically more
massive and hence more metal rich \citep{Tremonti:04:898} than the type I
LSNe, this would imply that WDML is the dominant source of the
circum-stellar matter and that the metals in the outer atmosphere have
their source in the host galaxy itself.  If, however, there is an intrinsic
source of metals in the stellar evolution of the progenitor from
atmospheric dredge-up or the material was ejected due to binary
interaction, then the SNe IIn-lum could be found in hosts of any mass.
Four of the brightest type I SN hosts in Figure~\ref{fig_cmd} have only
limits on $M_r$ and could move to fainter hosts leaving some of the
IIn-lums by themselves in brighter hosts.  Once again, only deeper
photometry and a larger sample will bring this relationship into focus.

\subsection{The $sSFR$ versus $M_*$ Diagram\label{sec_ssfr_vs_mstar}}

\begin{figure}
\includegraphics[angle=90., scale=0.40,viewport=24 54 528 608]{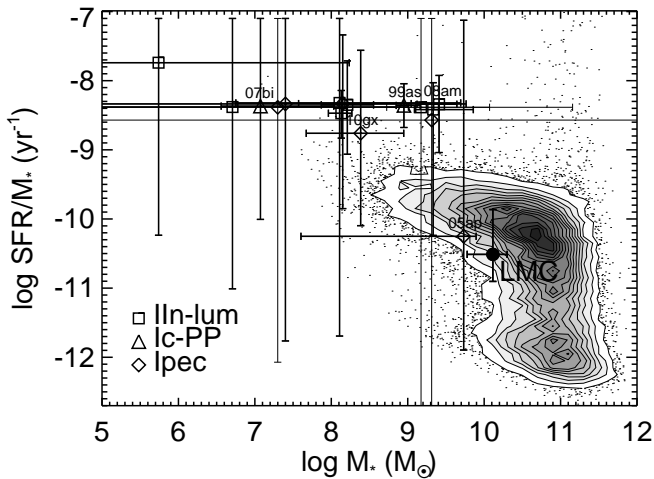}
\caption{Specific star formation rate as a function of stellar mass for the
LSN hosts with a selection of the LSNe labeled to avoid confusion due to
overlap.  The better constrained hosts have thicker error bars.
The contours represent the density of the larger sample of galaxies 
from the {\it GALEX}-SDSS cross match in \citet{Wyder:07:293} (see 
Figure~\ref{fig_cmd}).  The LMC is plotted for reference using values 
from \citet{Westerlund:97} and \citet{Harris:09:1243}.
\label{fig_ssfr}}
\end{figure}

Figure~\ref{fig_ssfr} illustrates the extreme nature of the LSNe hosts.
The error bars are large because in many cases our estimation of the host
SFH is derived from detection limits.  The grouping of the hosts near
$sSFR\sim-8.4$ is caused by the combination of a finite time-step in the
SFH models and the shortest lifetime of stars with SEDs that peak in the UV
($\sim10^8$ yr).  With smaller time-steps and age indicators more sensitive
to shorter timescales (e.g., H$\alpha$), the most likely values for $sSFR$
might be even higher.  Nearly every host is less massive and has a higher
$sSFR$ than the Large Magellanic Cloud.  The exception is SN2006gy (see
Table~\ref{tab_host_sed}), which appears in NGC1260, a peculiar S0/Sa
galaxy.  It has been pointed out that the site of SN2006gy is dusty
\citep{Ofek:07:L13,Miller:10:2218} and that the infra-red luminosity of the
host implies a SFR that is not inconsistent with the production of
high-mass stars \citep{Smith:07:1116}.  It is plausible that NGC1260 has
recently accreted a star-forming dwarf that is similar to the other LSN
hosts.

Given the low mass of the LSN hosts and their short sSFR timescales, it
appears that LSNe are produced in the infancy of a galaxy's evolution.  The
fact that SN2006gy is the only LSN host that appears in a high-mass host
also implies that LSNe are typically produced early in a galaxy's life,
before encounters with larger galaxies.  \citet{Young:10:70} point out that
SN2007bi presents a problem for PISN models that require either H-rich,
moderate metallicity progenitors ($Z\simeq Z_{\odot}/3$) or population III
objects with $Z\lesssim Z_{\odot}/1000$ \citep{Langer:07:L19}, because the
SN shows no evidence of being H-rich and yet the metallicity of the host is
not consistent with producing a population III progenitor
\citep[$12+\log({[O/H]})_{HOST} = 8.15\pm0.13$,][]{Young:10:70}.  It is
possible that the chemical evolution in such small dwarf galaxies is
heterogeneous and SN2007bi could have formed from a pocket of primordial
(pop III) gas.  \citet{Young:10:70} call for better metallicity
measurements of the host of SN2007bi, as their measurements were taken
while the SN continuum was still present.  It may also be important to
acquire resolved metallicity measurements to sort out the spatial pattern
of chemical enrichment in these young dwarf galaxies.

If we compare the LSN hosts with the supercompact UVLGs from
\citet{Hoopes:07:441}, we see that they share the same range of $sSFR$ and
have similar UV-optical colors.  The most massive LSN hosts are consistent
with the mass range for the supercompact UVLGs, but a large fraction of the
LSN hosts are less massive.  It is interesting to note that many of the LSN
hosts have masses similar to individual star-forming clumps in a sample of
LBAs measured in \citet{Overzier:09:203}.  This implies that LSN hosts may
delineate the low-mass tail of the LBAs, or that they are building blocks
from which LBAs and UVLGs are constructed through mergers.

\subsection{LSNe and The Stellar IMF\label{sec_imf}}

Finding stars as massive as any formed in the Milky Way in galaxies that
are many orders of magnitude smaller has strong implications for our
understanding of the upper end of the IMF.  It is exceedingly difficult to
measure individual stellar masses in any but the nearest galaxies, so the
LSNe offer the opportunity to test and calibrate theories of the IMF.

To illustrate the potential impact of LSNe on our understanding of the IMF,
we examine a recent theory based on the notion that the IMF in a given
galaxy is the result of integrating all the IMFs within individual clusters
in the galaxy, each of which has an IMF limited by the ongoing star formation
\citep{Kroupa:03:1076}.  This integrated galaxy intial mass function
(IGIMF) theory has successfully reproduced many observed properties of
high-mass star formation \citep{Altenburg:07:1550, Altenburg:09:394,
Altenburg:09:516, Weidner:10:275}.  One standard IGIMF scenario predicts a
relationship between the IGIMF and the ongoing SFR which is graphically
presented in Figure~4 of \citet{Altenburg:07:1550}.  This figure presents
IGIMF curves for a range of SFRs ranging from $10^{-5}$ to $10^2 M_{\odot}$
yr$^{-1}$.  If we compare the SFRs of the LSN hosts in
Table~\ref{tab_host_sed} with these curves, we see potential discrepancies
for some LSNe depending on what the initial masses are.  

We have an estimate for the initial mass of the progenitor of SN2007bi of
$>150 M_{\odot}$, and evidence that it was a single star, i.e., very little
circum-stellar material \citep{Yam:09:624}.  Our estimate of the SFR of the
host of SN2007bi ranges from $-1.97 < \log M_{\odot}$ yr$^{-1} < -0.48$
which is marginally consistent with the curves presented in Figure~4 of
\citet{Altenburg:07:1550} if the initial mass of SN2007bi is $150
M_{\odot}$ and not greater.  The most probable SFRs for many of the LSN
hosts are lower than the host of SN2007bi.  Discrepancies with the standard
scenario of the IGIMF theory could arise if any of the other LSNe have
progenitors with initial masses greater than SN2007bi.  We must, however,
remember that our SFRs for the LSN hosts are derived using methods that
assume a single universal IMF \citep{Kroupa:01:231,Sullivan:06:868}.  For
such small hosts, a single IMF may be appropriate, but this requires
spatially resolved imaging of the hosts to see if the majority of the star
formation is occuring in a single, large cluster.  A discrepancy may
indicate simply that the IGIMF curves need to be extended to higher stellar
masses.  A problem would exist then only if the production of such a high
mass star were exceptionally improbable.  This could be tested by
integrating the IGIMF to the mass of the LSN progenitor and comparing the
calculated total mass of the host to the observed mass.  Another refinement
of this comparison could be achieved by using SFRs based on indicators
sensitive to even shorter timescales, i.e., H$\alpha$ which is sensitive
over timescales of $\sim10^7$ yr.  The faintness of these hosts would
require a significant investment in observing time to achieve this.

On the other hand, any discrepancy could be evidence in favor of lower mass
progenitors, perhaps consistent with the magnetar scenario.  This
consistency for SN2007bi, narrow though it is, is potentially another
success for the IGIMF theory.  We can see, however, that definitive tests
await more accurate LSN progenitor mass estimates and a detailed
characterization of the star formation in the host galaxies.

\section{Conclusions\label{sec_conclude}}

The apparent preference that LSNe have for extreme host galaxies argues for
a local environmental effect in their production.  The mass-metallicity
relationship \citep{Tremonti:04:898} and the effect of metallicity on the
efficiency of stellar winds argues that WDML is the physical mechanism that
prevents LSNe from being produced in more normal, higher metallicity hosts.
The extreme nature of the LSN hosts is attested to by comparing their
distribution in Figures~\ref{fig_cmd} and \ref{fig_ssfr} with a much larger
sample of nearby galaxies from the SDSS - {\it GALEX} cross-match presented
in \cite{Wyder:07:293}.  Their distribution in $M_r$ may be different from
the general CC host distribution presented in \citet{Arcavi:10:777}, but a
measurement of this difference awaits deeper photometry and a larger sample
of LSNe hosts.  Measuring a difference between the hosts of type IIn and
type I LSN hosts also awaits better data.  The low SFR of the LSN hosts and
the possibly high initial mass estimates of the LSN progenitors places them
in a crucial location of theoretical diagrams relating the IMF or the IGIMF
to ongoing SFR.  Potential discrepancies with current theories may exist,
but only if typical host SFRs are less than $10^{-1} M_{\odot}$ yr$^{-1}$
and progenitor mass estimates significantly exceed $100 M_{\odot}$.



\acknowledgments

We greatefully acknowledge the anonymous referee for a careful reading and
useful suggestions that improved the presentation of this work.

Joint research by A.G. and M.S. is supported by the Weizmann-UK program.
A.G. is also supported by grants from the Israeli Science Foundation, an EU
FP7 Marie Curie IRG Fellowship, and a research grant from the Peter and
Patricia Gruber Awards.

GALEX (Galaxy Evolution Explorer) is a NASA Small Explorer, launched in
2003 April. We gratefully acknowledge NASA's support for construction,
operation, and science analysis for the GALEX mission, developed in
cooperation with the Centre National d'Etudes Spatiales of France and the
Korean Ministry of Science and Technology.

This research has made use of the NASA/IPAC Extragalactic Database (NED)
which is operated by the Jet Propulsion Laboratory, California Institute of
Technology, under contract with the National Aeronautics and Space
Administration.

Funding for the SDSS and SDSS-II has been provided by the Alfred P.
Sloan Foundation, the Participating Institutions, the National Science
Foundation, the U.S. Department of Energy, the National Aeronautics and
Space Administration, the Japanese Monbukagakusho, the Max Planck
Society, and the Higher Education Funding Council for England. The SDSS
Web Site is http://www.sdss.org/.

The SDSS is managed by the Astrophysical Research Consortium for
the Participating Institutions. The Participating Institutions are
the American Museum of Natural History, Astrophysical Institute
Potsdam, University of Basel, University of Cambridge, Case Western
Reserve University, University of Chicago, Drexel University,
Fermilab, the Institute for Advanced Study, the Japan Participation
Group, Johns Hopkins University, the Joint Institute for Nuclear
Astrophysics, the Kavli Institute for Particle Astrophysics and
Cosmology, the Korean Scientist Group, the Chinese Academy of
Sciences (LAMOST), Los Alamos National Laboratory, the
Max-Planck-Institute for Astronomy (MPIA), the Max-Planck-Institute
for Astrophysics (MPA), New Mexico State University, Ohio State
University, University of Pittsburgh, University of Portsmouth,
Princeton University, the United States Naval Observatory, and the
University of Washington.

The National Energy Research Scientific Computing Center, which is
supported by the Office of Science of the U.S. Department of Energy under
Contract No. DE-AC02-05CH11231, provided staff, computational resources and
data storage for this project.

\end{document}